\documentclass{aa}  

\pdfoutput=1

\usepackage{graphicx}
\usepackage{txfonts}
\usepackage{natbib,twoopt}
\usepackage[breaklinks=true]{hyperref} 
\bibpunct{(}{)}{;}{a}{}{,}             
\makeatletter
  \newcommandtwoopt{\citeads}[3][][]{\href{http://adsabs.harvard.edu/abs/#3}%
    {\def\hyper@linkstart##1##2{}%
     \let\hyper@linkend\@empty\citealp[#1][#2]{#3}}}
  \newcommandtwoopt{\citepads}[3][][]{\href{http://adsabs.harvard.edu/abs/#3}%
    {\def\hyper@linkstart##1##2{}%
     \let\hyper@linkend\@empty\citep[#1][#2]{#3}}}
  \newcommandtwoopt{\citetads}[3][][]{\href{http://adsabs.harvard.edu/abs/#3}%
    {\def\hyper@linkstart##1##2{}%
     \let\hyper@linkend\@empty\citet[#1][#2]{#3}}}
  \newcommandtwoopt{\citeyearads}[3][][]%
    {\href{http://adsabs.harvard.edu/abs/#3}
    {\def\hyper@linkstart##1##2{}%
     \let\hyper@linkend\@empty\citeyear[#1][#2]{#3}}}
\makeatother

\begin{document} 

   \title{TPCI: The PLUTO-CLOUDY Interface }

   \subtitle{A versatile coupled photoionization hydrodynamics code}

   \author{M. Salz\inst{1},
           R. Banerjee\inst{1},
           A. Mignone\inst{2},
           P. C. Schneider\inst{1},
           S. Czesla\inst{1},
           J. H. M. M. Schmitt\inst{1}
          }

   \institute{Hamburger Sternwarte, Universit\"at Hamburg,
               Gojenbergsweg 112, 21029 Hamburg, Germany\\
              \email{msalz@hs.uni-hamburg.de}
         \and
              Dipartimento di Fisica Generale, Universit\'a di Torino,
              via Pietro Giuria 1, 10125 Torino, Italy\\
             }

   \date{}


  \abstract{
            We present an interface between
            the (magneto-) hydrodynamics code PLUTO and the plasma simulation and spectral synthesis code CLOUDY.
            By combining these codes,
            we constructed a new photoionization hydrodynamics solver:
            {The PLUTO-CLOUDY Interface} (TPCI), which
            is well suited to simulate photoevaporative flows under strong
            irradiation.
            The code includes the electromagnetic spectrum from X-rays to the radio range and solves the photoionization and chemical network of the 30 lightest elements.
            TPCI follows an iterative numerical scheme: First,
            the equilibrium state of the medium is solved for a given radiation field by CLOUDY, resulting in a net radiative heating or cooling.
            In the second step, the latter
            influences the (magneto-) hydrodynamic evolution calculated by PLUTO.
            Here, we validated the one-dimensional version of the code on the basis of four test problems:
            Photoevaporation of a cool hydrogen cloud, cooling of coronal plasma,
            formation of a Str\"omgren sphere, and the evaporating atmosphere of
            a hot Jupiter.
            This combination of an equilibrium photoionization solver
            with a general MHD code provides an advanced simulation tool
            applicable to a variety of astrophysical problems.
   }

   \keywords{Methods: numerical,
             Hydrodynamics,
             Radiation: dynamics,
             Planets and satellites: atmospheres
             }

   \maketitle
%

\section{Introduction}

Hydrodynamic flows powered by strong irradiation (photoevaporation) can be found throughout the Universe from the evaporation of cosmological minihaloes \citep{Shapiro2004} and the formation and evolution of \ion{H}{II} regions \citep{Odell2001} to the evaporation of circumstellar disks \citep{Owen2010}. The discovery of expanding hot-Jupiter atmospheres \citep[][]{Vidal2003} has added another environment possibly dominated by photoevaporative mass loss \citep{Lammer2003}. Although photoevaporative phenomena occur on widely different spacial scales, the essential physical processes are similar: the absorption of ionizing radiation causes an increase in temperature and pressure, which results in a hydrodynamic wind.

Coupled photoionization and hydrodynamic simulations are essential for the progress of research in these fields. We created a new interface between an extensive equilibrium solver for the state of a gas or plasma under strong irradiation (CLOUDY) and a  3D (magneto-) hydrodynamics simulation code (PLUTO), the PLUTO-CLOUDY Interface (TPCI). The codes and the interface are publicly available\footnote{\href{http://www.nublado.org/}{http://www.nublado.org/}}\footnote{\href{http://plutocode.ph.unito.it/}{http://plutocode.ph.unito.it/}}\footnote{\href{http://www.hs.uni-hamburg.de/DE/Ins/Per/Salz/}{http://www.hs.uni-hamburg.de/DE/Ins/Per/Salz/}}.
The interface can be used to study photoevaporative processes under a wide range of conditions. However, we designed the code to study the environment of hot-Jupiter atmospheres, and a short introduction provides an idea of our requirements for the new simulation tool. 

The photoevaporation of hot-Jupiter atmospheres is the result of a complex interplay of various physical processes. The absorption of extreme ultraviolet radiation (EUV, $\lambda = 100 - 912$~\AA{})  heats the upper atmosphere of planets and in extreme cases triggers expansion and evaporation \citep{Watson1981}. Our focus lies on active host stars such as Corot-2 \citep{Schroeter2011}, where the effect of X-rays ($\lambda < 100$~\AA{}) on the mass-loss rate cannot be neglected \citep{Cecchi2006}. With standard abundances, metals dominate the absorption of X-rays \citep{Morrison1983} and  are in many cases important for the cooling of gases that
are devoid of molecules and dust \citep[e.g., in \ion{H}{II} regions, see][]{Osterbrock2006}. Thermal conduction can be a major heat source if strong temperature gradients occur in the atmospheres \citep{Watson1981}. Furthermore, hot Jupiters are most likely tidally locked, and eventually multidimensional simulations are needed to solve the disparate flows from the day and 
night 
side of the planets. Finally, planetary magnetic fields \citep{Trammell2011} or the interaction with the stellar wind \citep{Tremblin2013} also affect the evaporation process.

In this paper, we describe and validate the one-dimensional version of our numerical scheme without focusing on a particular science application. We start by describing our new coupled photoionization hydrodynamics simulation scheme (Sect.~\ref{secOverview}) and comparing it to available numerical codes (Sect.~\ref{secCodes}). 
The theoretical background and additional aspects of the interface are explained in Sects.~\ref{secPLUTO} to \ref{sec3Dsim}. 
To verify the results of the code and demonstrate the range of applications, four problems were drawn from different fields and were solved with TPCI (Sect.~\ref{secVerifi}). Our last test case is a simplified simulation of the escaping atmosphere of the hot Jupiter HD\,209458\,b, which we compare to previous studies of the same system (Sect.~\ref{secHJatmos}). In Sect.~\ref{secDiscus} we discuss the results and indicate future applications for TPCI.


\section{The PLUTO-CLOUDY Interface -- TPCI}
\label{secMethods}

\subsection{Overview}
\label{secOverview}

\begin{figure*}
  \centering
  \includegraphics[width=\hsize]{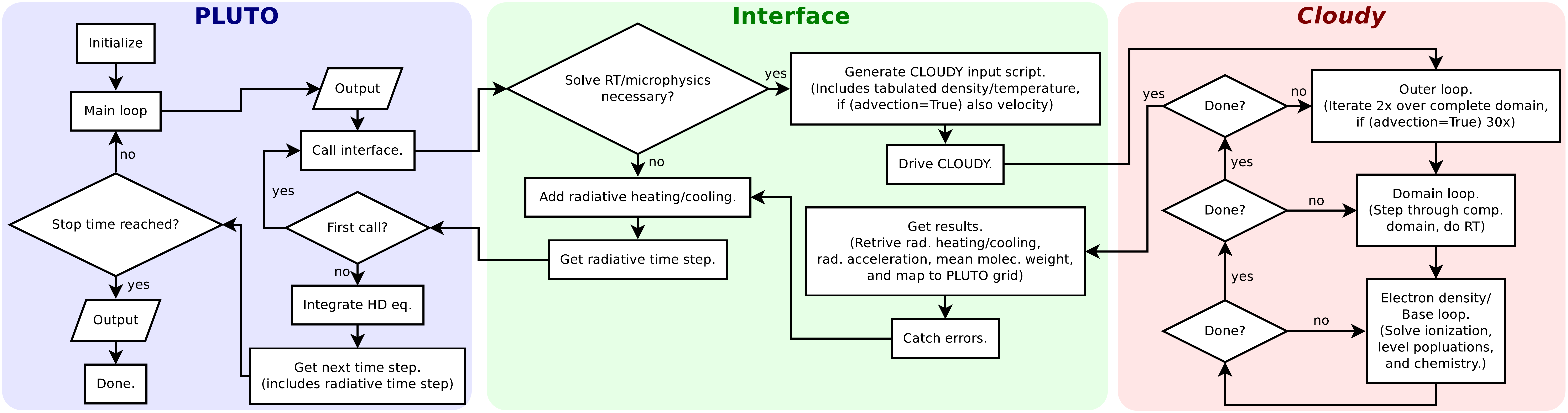}
  \caption{Flowchart of the PLUTO-CLOUDY Interface.
           Affiliation of modules is indicated by colored boxes.
           PLUTO solves the hydrodynamic evolution and CLOUDY
           the ionization and chemical equilibrium state of the gas
           under strong irradiation.
           Communication is achieved through the interface, which also handles the
           interpolation between the two different grid structures.
           }
      \label{FigTPCI}
\end{figure*}

We introduce an interface between two popular codes: the (magneto-) hydrodynamics code PLUTO \citep{Mignone2012} and the photoionization simulation code CLOUDY \citep{Ferland2013}. The interface allows studying steady-state and slowly evolving photoevaporative flows, in which the medium is in ionization and chemical equilibrium at all times during the simulation.

The numerical scheme follows a clear division between the solution of the (magneto-) hydrodynamic equations in PLUTO and the solution of the ionization and chemical equilibrium in CLOUDY (see Fig.~\ref{FigTPCI}). Basically, the photoionization solver is called by the hydrodynamic code with the density and temperature distributions and solves the microphysical state of the gas under the given conditions and with the specified irradiation. It returns the radiative heating or cooling distribution, which then affects the hydrodynamic evolution. The two steps are alternated during the simulation.

PLUTO is a freely distributed modular code for solving HD/MHD equations \citep{Mignone2007, Mignone2012}. It is a Godunov-type code that computes intercellular fluxes by solving the Riemann problem at cell interfaces. Parabolic terms in the differential equations such as viscosity or conductivity can be included by operator splitting, which is also the case for source terms such
as optically thin cooling or gravity. For TPCI we used the static grid version of the code.

CLOUDY is a photoionization and microphysics equilibrium solver for static structures, irradiated by an arbitrary source \citep{Ferland1998,Ferland2013}. It accounts for the electromagnetic spectrum from hard X-rays to the radio regime. The 30 lightest elements, from hydrogen to zinc, are included, and the solver balances radiative and collisional ionization and recombination, but also more complicated physical processes such as inner shell ionization or charge exchange\footnote{We sometimes us the generic term microphysics, meaning all processes that affect the equilibrium state in CLOUDY. For a full list we refer to the CLOUDY publications.}. In TPCI CLOUDY is called with a fixed temperature and density structure and solves the ionization and chemical equilibrium for the given conditions, which results in an imbalance of radiative heating and cooling. The program has been thoroughly tested to approach LTE at high densities and can be used within the temperature range from 3~K to 10$^{10}$~K and for number 
densities of up to 10$^{15}$~cm$^{-3}$. In cold regions CLOUDY approaches the fully molecular limit, including about $10^3$ chemical reactions mostly from the UMIST database \citep{
LeTeuff2000}. 
The ionization and chemical network is solved completely within CLOUDY, and only the mean molecular weight of the medium is passed to PLUTO. 
Since CLOUDY is a static equilibrium solver, advection of species is in general not included in TPCI. However, in 1D simulations with with a bulk flow antiparallel to the direction of irradiation, advective effects on the ionization equilibrium can be included via a CLOUDY internal iterative solver. Diffusion of elements is not included in this approach.

The coupled simulation codes use independent grid structures, and communication is achieved by linear interpolation on tabulated structures. TPCI is capable of performing true 3D MHD simulations coupled to pseudo-3D photoionization simulations along 1D parallel slices through the computational domain. One-dimensional simulations with the interface are serial, while 2D or 3D simulations are parallelized through the message passing interface (MPI). Here, we restrict the verification to 1D problems without magnetic fields. However, multidimensional simulations are not treated differently in the interface.

In the design of the interface we followed the paradigm to apply only minimal changes to both programs so as not to interfere with the well-tested numerical schemes while exploiting most of the capabilities of both codes. To this end, CLOUDY was used as an external library for solving the equilibrium state of the gas.
Several input and output extensions were necessary to enable the use of CLOUDY within TPCI: we adapted the code to accept tabulated density, temperature, and velocity structures, and the methods for retrieving the results from CLOUDY were expanded to include all necessary data in the communication with PLUTO.

The setup of TPCI follows the setup of the individual codes. While the usual input script for CLOUDY simulations is specified directly in the interface, the input files for the PLUTO code have not been changed. For the output of the hydrodynamic variables the standard PLUTO output options are available. Through the CLOUDY output much additional information is available, such
as the number densities of all ions and molecules, the important radiative heating and cooling agents with contribution to the total heating or cooling rate, and the irradiating, transmitted, and reflected spectra. In general, anybody who is familiar with the two simulation programs individually will be able to use the interface with only little additional introduction.

\subsection{Other photoionization hydrodynamics simulation codes}
\label{secCodes}

For reference, we provide a short overview of some available radiation hydrodynamics codes and compare individual aspects of the numerical schemes with our new interface.

For instance, \citet{Owen2010} coupled the 3D photoionization and radiative transfer code MOCASSIN \citep{Ercolano2003} with the hydrodynamic code ZEUS-2D \citep{Stone1992}. MOCASSIN and CLOUDY are similarly extensive equilibrium photoionization solvers, but the temperature parameterization used by \citeauthor{Owen2010} is only valid for X-ray heating.

Another example for coupling a microphysical equilibrium solver to a hydrodynamic simulation is the ionization module for the FLASH code \citep{Fryxell2000} presented by \citet{Rijkhorst2006} and further improved by \citet{Peters2010}. The 3D radiative transfer method is highly efficient in simulations with adaptive mesh refinement on distributed systems, but is computationally more demanding than the pseudo-3D scheme used here. A similarly advanced parallel radiative transfer method was introduced by \citet{Wise2011} into the ENZO code \citep{Bryan1997, OShea2004}; it is called MORRAY. The non-equilibrium chemistry solver is restricted to hydrogen and helium, however \citep{ENZO2013}.

\citet{Shapiro2004} and predating publications have extended the hydrodynamics code CORAL \citep{Raga1995} to include radiative transfer and non-equilibrium photoionization of hydrogen, helium, and metals. The scheme uses a similar pseudo-multidimensional radiative transfer method, but neglects X-rays, which is one of our main interests.
X-rays are also mostly neglected in the numerical schemes, which have been specifically designed to simulate escaping hot-Jupiter atmospheres, and the authors focus exclusively on 1D simulations \citep[e.g.,][]{Yelle2004, Tian2005, Garcia2007, Penz2008-2, Murray2009, Koskinen2013}.

In comparison, only TPCI solves our need for a photoionization hydrodynamics solver including hydrogen, helium, and metals as well as the absorption of EUV and X-ray emission.

\subsection{Hydrodynamic simulation --- PLUTO}
\label{secPLUTO}

In the following we present only the equations that are most
relevant for the theoretical background of TPCI. The hydrodynamic simulation part is solved by the code PLUTO \citep{Mignone2007, Mignone2012}.
For a single fluid with mass density $\rho$, pressure $p,$ and velocity $\mathbf{v}$, the hydrodynamic equations in the quasi-conservative form are given by
\begin{equation}\label{eqHD}
  \frac{\partial \mathbf{U}}{\partial t}
  + \nabla\cdot \mathrm{F}
  = \nabla\cdot \Pi
  + \mathbf{S_p} \,.
\end{equation}
Here $\mathbf{U}=(\rho, \rho\mathbf{v}, \mathcal{E})$ denotes the vector of conserved variables: mass, momentum, and total energy density, which is given by $\mathcal{E} = p/(\Gamma -1) + 1/2\rho v^2 $, where $\Gamma$~($= 5/3$) is the specific heat ratio of a monatomic gas. $\mathrm{F}=(\rho\mathbf{v}, \rho\mathbf{vv} + p, (\mathcal{E}+p)\mathbf{v})$ is the hyperbolic flux tensor, and $\Pi$ is the parabolic flux tensor, which in our case only
contains the conductive energy flux $\mathbf{F_c}$. $\mathbf{S_p}$ represents source terms such as gravity  $\mathbf{S_G} = \left(0,\rho \mathbf{a_{G}},\rho \mathbf{v}\cdot\mathbf{a_{G}}\right)$ with the gravitational acceleration $\mathbf{a_{G}}$, and the radiative source term $\mathbf{S_R} = \left(0,0,(G_{\mathrm{R}} - L_{\mathrm{R}})\right)$, where $G_{\mathrm{R}}$ and $L_{\mathrm{R}}$ are the radiative heating and cooling rates.

We follow a standard  operator split scheme and include the radiative heating or cooling rate between the hydrodynamic integration steps. First, the homogeneous left-hand side of Eq.~\ref{eqHD}, namely the Euler equations, are solved. Second, PLUTO computes the included parabolic and source terms. Third, CLOUDY is called to compute 
the radiative source term (see Sect.~\ref{secCloudy}), which subsequently affects the internal energy through the thermal pressure:
\begin{equation}\label{eqRadSource}
  p^{n+1} = p^{n} + \Delta t^{n} (\Gamma -1)S^{n}_{\mathrm{R}} \,,
\end{equation}
where $\Delta t^{n}$ denotes the explicit time step at integration step $n$.

To minimize the computational effort, the radiative source term is not updated at every time step, but only if a change in either pressure or density by more than a user defined {\tt change factor} triggers a call to the photoionization solver. A value of 1\% for the {\tt change factor} is a good compromise between accuracy and convergence speed for an initial settling phase of steady-state solutions. The accuracy can be increased by restarting the code with a smaller {\tt change factor}.

\subsection{Photoionization solver --- CLOUDY}
\label{secCloudy}

TPCI uses CLOUDY to solve the photoionization and the equilibrium state of the medium.
We provide a compact overview of the most important processes for our applications in CLOUDY; for a more general introduction see \citet{Ferland1998, Ferland2013}. The code steps through a one-dimensional slice of the computational domain with an adaptive step width using linear interpolation on the tabulated density and temperature structures passed by PLUTO. In each cell it solves the local equilibrium state by taking into account ionization and recombination processes, chemical reactions, and atomic level transitions of all included elements.

CLOUDY separates the radiative transfer of continuum and line radiation. The continuum is diminished by absorption,
\begin{equation}\label{eqContAbs}
  I = I_0 \exp \left( -\mathrm{d}\tau_{\mathrm{abs}} \right)
,\end{equation}
where $I$ is the intensity after the initial intensity $I_0$ passes a slab of gas with the total optical depth $\mathrm{d}\tau_{\mathrm{abs}}$. The total optical depth contains all continuous opacity sources included in CLOUDY, but in the EUV range it is dominated by ionization processes of neutral hydrogen.
The continuum is furthermore diminished by scattering \citep{Lightman1988}:
\begin{equation}\label{eqContScat}
  I = I_0 \left(1 + \frac{1}{2}\,\mathrm{d}\tau_{\mathrm{scat}} \right)^{-1} \,.
\end{equation}
The total scattering optical depth $\tau_{\mathrm{scat}}$ also
contains the opacities of several processes, of which Rayleigh scattering in the Lyman line wings dominates in the UV range.

The escape probability mechanism \citep[][]{Castor1970, Elitzur1982} is used to approximate radiative transfer effects. The local mean intensity $J$, needed to solve the rate equations, is then given by
\begin{equation}\label{eqEscProb}
  J =  S \left( 1-P_{\mathrm{esc}} \right)  \,.
\end{equation}
Here $S$ is the source function and $P_{\mathrm{esc}}$ the escape probability, which only depends on the optical depth. Solving the local equilibrium state is thereby decoupled from the radiative transfer equation. This approximation is exact in a single cell, where the source function is constant \citep{Elitzur1992}. CLOUDY ensures this by choosing the adaptive grid resolution so that neither density, temperature, heating, or other properties change strongly within one cell. However, it is necessary to step through the computational domain at least twice to obtain a reasonable estimate for the optical depth in the computation of the escape probability. This iteration is needed because radiation can escape in both directions, that is, away and toward the source.

CLOUDY can include the 30 lightest elements with any abundances. These species are only present in the photoionization solver, only the mean molecular weight is passed to the hydrodynamic single fluid simulation. The ionization ladder of every element is computed by balancing the rate equations
\begin{equation}\label{eqIonRate}
  \frac{\partial n(X^i)}{\partial t}
  = n_{\mathrm{e}}n(X^{i+1})\alpha(X^{i},T) 
    - n(X^i) \left(\Gamma(X^i) + n_{\mathrm{e}}\alpha(X^{i-1},T) \right)
  = 0
\end{equation} 
for each ionization stage. Here $n(X^i)$ is the number density of the element $X$ in the ionization stage $i$, $n_{\mathrm{e}}$ is the electron number density,
$\alpha(X^{i},T)$ is the temperature-dependent recombination rate from $X^{i+1}$ to $X^i$ summed over all levels, and $\Gamma(X^i)$ is the photoionization rate given by
\begin{equation}\label{eqPhotIonizRate}
  \Gamma(X^i) = \int_{\nu_{0,i}}^{\infty} \frac{4\pi J_{\nu}}{h\nu} a_{\nu}(X^i) \,\mathrm{d}\nu \,.
\end{equation} 
Here $J_{\nu}$ is the mean intensity at frequency $\nu$, $a_{\nu}(X^i)$ is the photoionization cross-section, and $\nu_{0,i}$ is the threshold frequency above which photoionization is possible. For simplicity we do not show more complex physical processes such as secondary ionizations or charge exchange in Eq.~\ref{eqIonRate}, which are also solved by CLOUDY, however.

For the hydrodynamic part of the simulation we need the radiative heating or cooling rates. While the heating rate $G_{\mathrm{R}}$ is a sum of many processes, the main heat source is usually given by photoionization:
\begin{equation}\label{eqIonizHeatRate}
  G_{\Gamma} = \sum_{X^i} n(X^i) \int_{\nu_{0,i}}^{\infty} \frac{4\pi J_{\nu}}{h\nu}
                                 h\left(\nu-\nu_{0,i}\right) a_{\nu}(X^i) \,\mathrm{d}\nu \,;
\end{equation}
the sum includes all species. Locally other heat sources such
as line absorption, charge exchange, and chemical reactions can dominate the heating rate.

Many processes contribute to the cooling rate $L_{\mathrm{R}}$ of a gas: recombination processes, free-free emission of electrons in the field of ions, line radiation, free-bound transition of H$^{-}$ , and chemical reactions, to name a few. Cooling due to recombinations is given by
\begin{equation}\label{eqRecombCoolRate}
  L_{\mathrm{L}} = n_{\mathrm{e}}n(X^{i+1}) kT \beta(X^{i},T) P_{\mathrm{esc}} \,,
\end{equation}
where $\beta(X^{i},T)$ stands for the effective recombination coefficient, averaged over the kinetic energy of the electron gas. The escape probability $P_{\mathrm{esc}}$ ensures that recombinations only contribute to the cooling rate if the emitted photon escapes the medium.
Another main cooling agent is line radiation of collisionaly excited elements:
\begin{equation}\label{eqLineCoolRate}
  L_{\mathrm{L}} = n_{\mathrm{e}}\left( n_{\mathrm{l}}c_{\mathrm{lu}} -n_{\mathrm{u}}c_{\mathrm{ul}} \right) h\nu_{\mathrm{lu}} \,.
\end{equation}
Here $c_{\mathrm{lu}}$ denotes the collision rate between the lower and upper levels of the atom. The difference of collisional excitation and de-excitation is emitted radiatively and leads to cooling. The level populations of each ionization stage of every element are solved by balancing the equilibrium radiative and collisional transition rates in model atoms of differing complexity. Thus, CLOUDY includes a large number of lines \citep[$10^5 -10^6$,][]{Ferland2013}.

\subsection{Time step control and relevant timescales}

In TPCI simulations, restrictions due to hydrodynamic and microphysical timescales must be observed.
PLUTO chooses the next hydrodynamic time step based on the Courant-Friedrichs-Levy (CFL) condition \citep{Courant1928}, which basically restricts disturbances in the medium to propagate by less than one cell width per time step. Radiative heating and cooling additionally limits the hydrodynamic time step in TPCI, meaning that the energy loss due to radiative cooling cannot exceed the internal energy. It is efficient to limit the next time step depending on the fraction of the energy content and the radiative source term \citep{Frank1994}:
\begin{equation}\label{eqDTRad}
  \Delta t^{n+1}_{\mathrm{TPCI}} =  \epsilon_{\mathrm{R}} 
                       \frac{p^{n}}{(\Gamma -1)S^{n}_{\mathrm{R}}}\,,
\end{equation}
and the next hydrodynamic time step is then given by
\begin{equation}\label{eqDTmin}
  \Delta t^{n+1} = \mathrm{min}
                   \left[\Delta t^{n+1}_{\mathrm{PLUTO}},
                   \Delta t^{n+1}_{\mathrm{TPCI}} \right] \,.
\end{equation}
Here $\epsilon_{\mathrm{R}}$ is the user-defined maximum fractional variation parameter. The default value is $\epsilon_{\mathrm{R}} = 0.1$, which restricts the change in internal energy due to radiative heating or cooling to be lower than 10\% in every time step.

CLOUDY solves the equilibrium state of the medium based on a diversity of processes, each with a characteristic timescale, and the use of TPCI implies that the microphysical timescales are shorter than the hydrodynamic timescale.
In gases with a temperature of about 10\,000~K, hydrogen recombination usually is the microphysical process with the longest timescale \citep{Ferland1979}:
\begin{equation}\label{eqTimeRec}
  T_{\mathrm{rec}} = \frac{1}{\alpha_{\mathrm{A}}(T_{\mathrm{e}})n_{\mathrm{e}}}
                   \approx 1.5\times10^{9} \, T_{\mathrm{e}}^{0.8} n_{\mathrm{e}}^{-1} \,.
\end{equation}
TPCI locally checks whether recombination is slower than advection and issues a warning. Equation~\ref{eqTimeRec} holds only for ionized regions; especially molecular reactions occur on longer timescales. The CLOUDY output provides a manual method to control the longest timescales and, thus, check the validity of the approach.

In 1D simulations, the advection of species can be included through a steady-state solver in CLOUDY (see Sect.~\ref{secTPCI_Advec}). With this scheme, steady flows can be simulated in which the advective timescales are shorter than microphysical timescales.
Nevertheless, the assumption of photoionization and chemical equilibrium is a strong simplification, and the validity must be considered for every application.

\subsection{Advection of species}
\label{secTPCI_Advec}

The impact of advection due to a bulk flow of the medium can be included in the simulation by using an iterative steady state solver included in CLOUDY \citep{Henney2005}. The solver is available for simulations in which the medium flows toward the source of ionizing radiation. The solution of the equilibrium state then includes source and sink terms of all species, induced by material flows:
\begin{equation}\label{eqAdvection}
  \nabla\cdot\left(n(X^i)\mathbf{v}\right) = n\mathbf{v}\cdot\nabla\left(n(X^i)/n\right) \,.
\end{equation}
In this case, PLUTO passes the velocity in addition to the density and temperature to the photoionization solver. CLOUDY computes the advective source terms using the relative density \citep{Henney2005}, which is an advective scalar.

The steady-state solver in CLOUDY includes advection in an iterative manner, and the user must control the convergence of the solution. The module defines the advection length $\Delta z$, a measure for the resolution of advection processes. It is refined if the advective solution has converged for the current advection length. The progress is controlled by following the discretization error and the convergence error \citep[see][for a detailed description]{Henney2005}. The convergence error is given by
\begin{equation}\label{eqConvError}
  \epsilon^2_1 =  \sum_{X^i, j}\left(\frac{n_j^{m}(X^i) - n_j^{m-1}(X^i)}{\Delta z/v}\right)^2 \, ,
\end{equation}
where $m$ is the iteration number and j is the zone number; the sum extends over all elements and ionization stages as well as over all zones. The error approaches zero as the changes in number densities of consecutive iterations decreases. The discretization error is defined as
\begin{equation}\label{eqDiscretError}
  \epsilon^2_2 =  \sum_{X^i, j}\left(\frac{n_j^{m}(X^i) - n_{(j-\Delta z)}^{m-1}(X^i)}{\Delta z/v}
                                   - \frac{n_j^{m}(X^i) - n_{(j-0.5\Delta z)}^{m-1}(X^i)}{\Delta z/2v} \right)^2 \, .
\end{equation}
$(j-\Delta z)$ is a symbolic index referencing the density of the species at the distance of the advection length. The error describes the differences of the local, advective source or sink terms to the value if the advection length is refined by a factor of two. The sum again runs over all species and zones.

The discretization error depends on the resolution in the photoionization solver. This resolution can be up to ten times higher than in the hydrodynamic part of the simulation. For TPCI it is more important to check the advection length than the actual value of the discretization error. The advection length is refined if  $\epsilon^2_1 < 0.1 \epsilon^2_2$. It should generally be decreased to the hydrodynamic grid spacing, so that advective effects have the same accuracy as the rest of the simulation. 

The described procedure does not follow our split scheme to solve hydrodynamic effects exclusively in PLUTO. However, at the moment we cannot solve the  advection of species in PLUTO. It is possible to include advective scalars for the relative density of all species in PLUTO, but then CLOUDY would have to be called with the densities of all 30 elements and ionization stages, which is not implemented. The steady-state solver in CLOUDY is computationally demanding. One advection solution may need up to 100 iterations to converge because advective effects must be swept through the computational domain \citep{Henney2005}. Thus, including advection slows simulations down, and when searching for advective steady-state solutions, it is advisable to use the non-advective code for an initial settling of the hydrodynamic structure and then restart the simulation including advection to let the structure evolve to the final steady state. This procedure has been followed in the two test problems, which include advection 
of species (see Sects.~\ref{secWeakD}~and~\ref{secHJatmos}).

\subsection{Radiative acceleration}
\label{secRadAccel}

In addition to affecting the temperature of the gas, the absorption of radiation also exerts a pressure on the medium. CLOUDY records the radiative pressure produced by the absorption of radiation from the central source. The local radiative acceleration can be directly accessed after a CLOUDY call. This acceleration is handled like any external force by the PLUTO code (see Eq.~\ref{eqHD}).

\subsection{Pseudo-multidimensional simulations}
\label{sec3Dsim}

Although here we focus on 1D simulations,
TPCI can be used in two or three dimensions on a Cartesian grid with irradiation along the x-axis. The radiative solution is then simply split into parallel 1D simulations, which is called a pseudo-3D radiative transfer. There is no true 3D radiative transfer in CLOUDY, but the coupling of 3D radiative transfer to hydrodynamics simulations is in many cases still computationally prohibitive today \citep[e.g.,][]{Owen2010}. In problems with a strong directionality in the radiative field, the pseudo-3D approximation is valid, unless there are shadow-casting objects \citep{Morisset2005}. However, in the case of strong shadows, the induced error can also be neglected if the energy transfer into the shadow region is hydrodynamically dominated.

The 1D version of the interface is a serial code because CLOUDY requires the complete density structure to solve the plane-parallel radiative transfer through the computational domain. PLUTO is, however, capable of 3D parallel computations, which is also enabled in TPCI by not decomposing the domain along the x-axis. Each thread then deals with the full 1D structure along the x-axis for a given number of y/z-grid-points. After the complete domain has been processed, the hydrodynamic evolution continues until the next call to CLOUDY is necessary. For this approach the computational domain should be decomposed so that the  available processors have the same number of y/z-points to minimize idle times.

This pseudo-3D scheme provides a convenient possibility to allow parallel simulations with low complexity.
In the best case, a multidimensional simulation will have a similar execution time as a 1D simulation if every y/z-grid-point uses an individual processor. The communication overhead is small, since the main computational load usually comes from the independent CLOUDY calls.


\section{Verification}
\label{secVerifi}

In the following we present TPCI simulations of four test problems.
The first two simulations verify the
implementation of our numerical scheme, each with a different focus.
The third test problem is included to clarify the limits of validity of TPCI.
The fourth test case is a simplified simulation of a hot-Jupiter atmosphere,
which already demonstrates the capabilities of TPCI in this field. 
The setup of each simulation is given in Table~\ref{tabSim}.

\begin{table*}
\caption{Simulations for the verification of TPCI}             
\label{tabSim}      
\centering          
\begin{tabular}{l l c@{\,}c@{\,}c c@{\,}c@{\,}c l@{\,\,\,\,\,}l l@{\,\,\,\,\,}l }
\hline\hline\vspace{-9pt}\\
  Name     & Numerical setup                         & \multicolumn{6}{l}{BC}                               & \multicolumn{2}{l}{Grid}                   & \multicolumn{2}{l}{Irradiation} \\ 
           & {\small (time stepping, interpolation,} & \multicolumn{3}{c}{left} & \multicolumn{3}{c}{right} & {\small geometry} & {\small  grid points}  & {\small SED shape} & {\small ioniz. radiation field} \\ 
           & {\small  Riemann solver)}                & $\rho$ & $v$ & $p$ & $\rho$ & $v$ & $p$              &                   &                        &                    & {\small $(\lambda<912~\AA{})$}\\ 
\vspace{-9pt}\\ \hline\vspace{-9pt}\\              
  D-front  & {\tt RK3, WENO3, $\,\,\,$hllc}          & f & f & o & o & f & o                                & Cartesian & 200 uni.                      & 50\,000~K BB        & $\Phi_{\mathrm{H}} = 10^{11}$~cm$^{-2}$\,s$^{-1}$ \\  
           & {\tt +advection}                        &   &   &   &   &   &                                  &           &                               &                     & \\
  c-shock  & {\tt RK3, LINEAR, two\_shock}           & o & o & o & o & o & o                                & Cartesian & 400 uni.                      & ---                 & --- \\  
  R-front  & {\tt RK3, LINEAR, hllc}                 & o & o & o & o & o & o                                & Cartesian & 400 uni.                      & 50\,000~K BB        & $Q_{\mathrm{H}} = 10^{49}$~s$^{-1}$ \\ 
  hJupiter & {\tt RK3, WENO3, $\,\,\,$hllc}          & f & o & f & o & o & o                                & Spherical & 240 stretch.                   & solar min.          & $4\pi J = 1315$~erg\,cm$^{-2}$\,s$^{-1}$ \\   
           & {\tt +gravity, +thermal cond.}          &   &   &   &   &   &                                  &           &                               &                     & \\
           & {\tt +advection}                        &   &   &   &   &   &                                  &           &                               &                     & \\
\vspace{-9pt}\\ \hline                  
\end{tabular}
\tablefoot{Abbreviations: (BC) boundary condition,
           (f) fixed boundary condition,
           (o) outflow boundary condition,
           (SED) spectral energy distribution,
           (uni.) uniform grid spacing,
           (stretch.) stretched grid spacing,
           (BB) blackbody spectrum,
           (solar min.) solar minimum spectrum \citep{Woods2002},
            $\Phi_{\mathrm{H}} = Q_{\mathrm{H}}/4\pi r_0^2$,
            $Q_{\mathrm{H}} = \int_{\nu_0}^{\infty} L_{\nu}/h\nu \,\mathrm{d}\nu$,
            $4\pi J = \int_{\nu_0}^{\infty} 4\pi J_{\nu}\,\mathrm{d}\nu$,
           ({\tt RK3}) third-order Runge-Kutta scheme \citep{Gottlieb1996},
           ({\tt WENO3}) weighted essentially non-oscillatory finite difference scheme \citep{Jiang1996},
           ({\tt hllc}) Harten, Lax and van Leer approximate Riemann solver with contact discontinuity \citep{Toro1994},
           ({\tt two\_shock}) nonlinear Riemann solver based on the two-shock approximation \citep{Colella1984, Fryxell2000}.
           }
\end{table*}

\subsection{Weak D-type ionization front}
\label{secWeakD}

For the initial verification, we investigated the effects of
steady-state flows in 1D simulations of ionization
fronts.
Such flows appear when molecular clouds are evaporated by the ionizing
emission of newly formed stellar clusters.
The gas is heated through ionization processes, and the
cloud slowly evaporates.
If the evaporation proceeds as a
subsonic wind from a dense region, it is called a weak D-type ionization front,
in contrast to strong ionization fronts, which
involve transonic flows.
(For rarefied R-type ionization fronts see Sect.~\ref{secRtype}.)

The results from our TPCI simulation
are compared with an independent CLOUDY simulation, using
the included steady-state solver, which was specifically designed to solve these photoevaporative
flows (see \citet{Henney2005})\footnote{The CLOUDY internal, iterative, steady-state hydrodynamics solver is restricted to specific flows and geometries. TPCI is more versatile and solves the dynamic evolution of the studied flows.}.
This is a powerful test for TPCI because both simulations rely on the same
photoionization solver, thus, any differences in the results can only be
caused by the interaction of the hydrodynamics and the photoionization
solver through the interface.

\begin{figure}
\centering
\includegraphics[width=0.95\hsize]{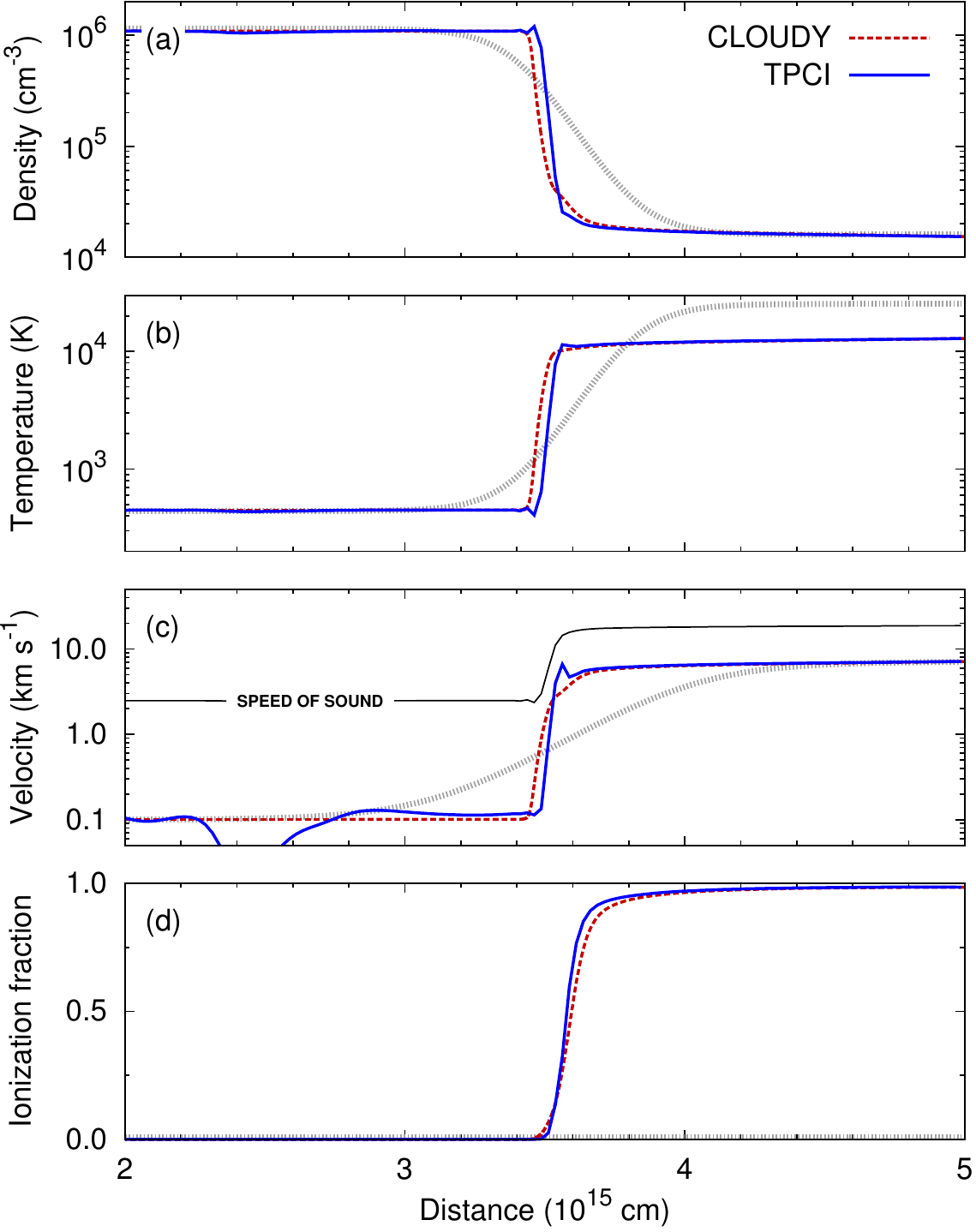}
   \caption{Pure hydrogen weak D-type ionization front.
            Density (a), temperature (b), velocity (c), and
            degree of ionization (d) are depicted.
            A CLOUDY simulation (red dashed lines) is compared with
            a TPCI simulation (blue solid lines).
            Initial conditions are given by the gray dotted lines.
            In panel (c) the speed of sound is shown by the
            black solid line.
            The TPCI simulation 
            agrees well with the independent CLOUDY simulation.
            Small oscillations in the high-density region that can be
            seen in panel (c) do not affect the solution significantly.
           }
      \label{FigReproHenney}
\end{figure}

Figure~\ref{FigReproHenney} shows the simulation of a weak D-type
ionization front in a pure hydrogen gas in the rest frame of the front
together with the initial conditions.
Cold, neutral gas (400~K) with a density of $10^{6}$~cm$^{-3}$ 
is irradiated from the right-hand side by a hot O-star
with a 50\,000~K blackbody spectrum.
The gas shows a steady-state flow toward the source of
ionizing radiation.
The evaporation process is subsonic, starting with
0.1~km\,s$^{-1}$ and reaching 7.1~km\,s$^{-1}$ behind
the ionization front.
The temperature increases to 12\,000~K and the
density drops to $2\times10^{4}$~cm$^{-3}$.

\begin{figure}
\centering
\includegraphics[width=\hsize]{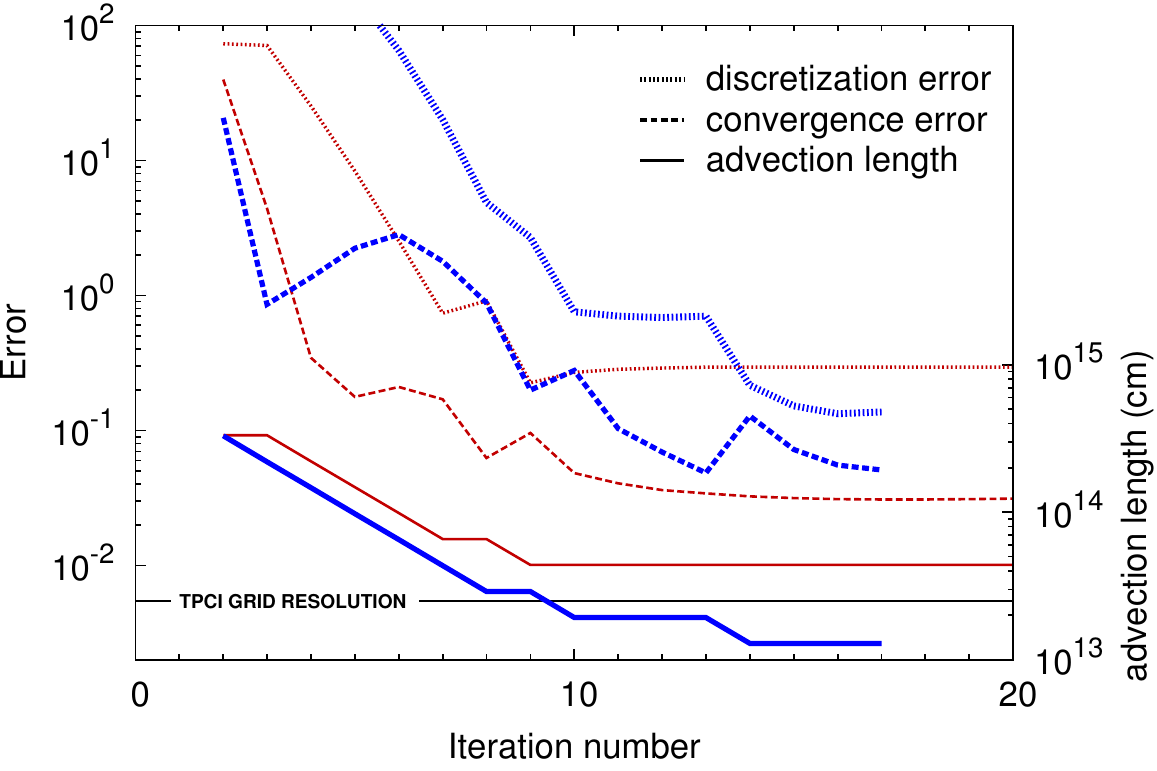}
   \caption{Convergence of the models of the weak D-type 
            ionization front.
            The CLOUDY simulation is shown
            by red, thin lines, the advective TPCI simulation
            by thick, blue lines.
            The discretization error (dotted), the convergence
            error (dashed), and the advection length (solid) are
            plotted against the iteration number of the steady-state 
            solver in CLOUDY.
            The grid resolution of the TPCI simulation is indicated
            by the thin, black line.
            In both simulations the advection length is reduced to within
            a factor of 2 of the grid resolution in the hydrodynamic
            simulation part.
           }
      \label{FigReproHenneyConv}
\end{figure}

We used TPCI without the advection of species
for the initial settling of the simulation and subsequently restarted
the simulation including the advection as described in Sect.~\ref{secTPCI_Advec}.
The settling phase of the dynamical simulation took $3\times10^{4}$~a,
after which oscillations of the ionization front
were reduced to about 1\% of the depth of the
ionized region.
These oscillations in the denser gas
are caused by the reflection of waves at the
left-hand boundary and at the ionization front.
The depth of the ionized region in the steady-state CLOUDY simulation
agrees with the \emph{dynamic} TPCI simulation within the
errors produced by the remaining oscillations
(see Fig.~\ref{FigReproHenney}~(d)).
The width of the ionization front is reduced by 25\% in the dynamic
simulation, which does not affect the overall structure, however.

The convergence of the steady-state solver in the advective TPCI and the CLOUDY
simulations is shown in Fig.~\ref{FigReproHenneyConv}.
The advection length has been iteratively refined whenever
the solution converged for the current length 
(see Sect.~\ref{secTPCI_Advec}).
Every refinement temporarily increases the convergence error,
which is then reduced over the following iterations until
further refinement.
The final discretization error is on the order of 10\%,
computed on the fine grid of the photoionization solver.
This is sufficient because the advection length is similar
to the coarser resolution of the hydrodynamical simulation part.
Higher refinement is impossible because in this case the steady-state solver
does not proceed beyond this level (see Sect.~\ref{secDiscus} for
a discussion of this problem).

The comparison of the two simulations demonstrates that the
interface works correctly and is capable of simulating
photoevaporative flows.
While the independent CLOUDY simulation with the internal steady-state
solver produces smoother profiles, TPCI allows studying the dynamical evolution
of the phenomenon.

\subsection{Coronal flare - chromospheric evaporation wave}
\label{secCoronalFlare}

\begin{figure}
\centering
\includegraphics[width=\hsize]{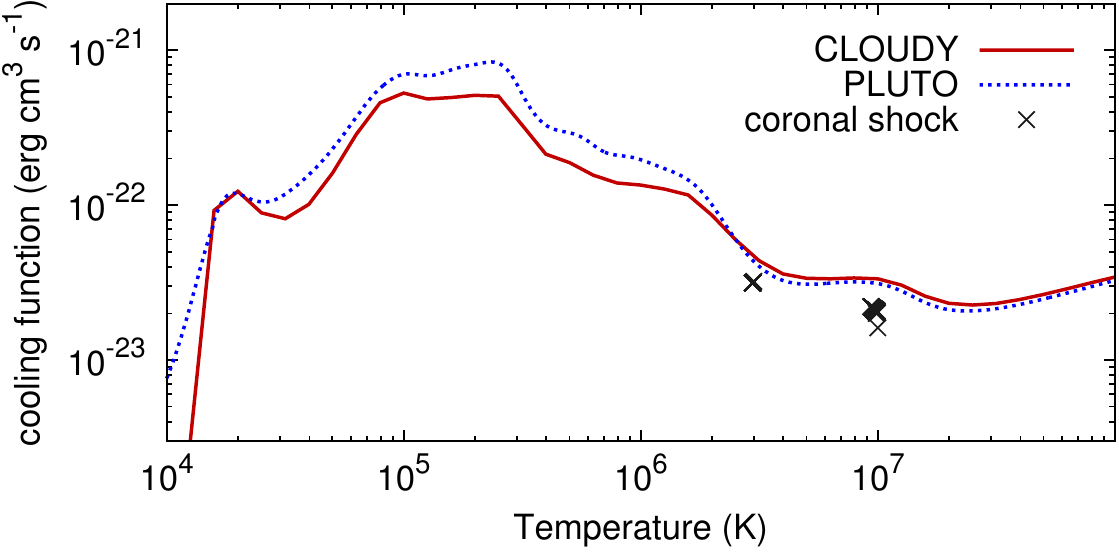}
   \caption{Optically thin cooling function of PLUTO (blue, dotted) 
            and CLOUDY (red, solid).
            The crosses depict the
            actual cooling in the coronal shock simulation with TPCI.
            These are a factor of 1.5 lower because the simulation shows 
            significant optical-depth effects.
           }
      \label{FigCoolingFuncs}
\end{figure}

TPCI can also make use of CLOUDY to compute the cooling
of collisional plasmas instead of strongly
irradiated plasmas.
This provides the possibility of testing the interface 
in the optically thin limit, where
radiative losses only depend on the electron number
density $n_\mathrm{e}$ and the cooling function
$\Lambda_\mathrm{R}$ \citep[e.g., ][]{Sutherland1993}:
\begin{equation}\label{EqCoolOpThin}
  L_\mathrm{R} = n_{\mathrm{e}}^2\Lambda_\mathrm{R}(T,A)\,.
\end{equation}
The cooling function depends on temperature $T$ and metal
abundance $A$.
A radiative transfer solver is not necessary 
for the simulation as long as the optical depth is negligible.
Cooling in this regime is dominated by emission from ionized metals,
in which case CLOUDY mostly uses transitions from the CHIANTI
database \citep{Dere1997, Landi2012}.
The resulting cooling function with solar abundances in CLOUDY is similar to standard
optically thin cooling functions; see Fig.~\ref{FigCoolingFuncs}
for a comparison with the tabulated cooling function that is included
in the PLUTO code.

We can thus compare a TPCI simulation with the radiative transfer and
microphysics solver with a PLUTO simulation using the included cooling function.
In contrast to the ionization front simulation in the previous section, we now used the same hydrodynamics
solver (i.e., PLUTO), but different solvers for the radiative cooling. 
Differences in the results of the test simulations can only result from the
implementation of the radiative cooling from CLOUDY in PLUTO. 
In this simulation we included all 30 elements available in CLOUDY, which additionally
tests the implementation of metals in the inteface.

We used a setup that was adapted to a strong
chromospheric evaporation wave in a 1D magnetically confined
coronal loop.
Such evaporation fronts appear when a magnetic reconnection
event releases a large amount of energy in the solar
atmosphere, which leads to a solar flare \citep[e.g.,][]{Reale2008}.
Evaporation of chromospheric plasma into the corona
causes a wave that propagates at speeds of
$\sim 400$~km\,s$^{-1}$ through a coronal loop,
increasing the density by up to one order of magnitude.
Since optically thin cooling depends on the
density squared, the cooling rate is increased
and, in the long term, the evaporated plasma
cools down and falls back onto the chromosphere.
The magnetic loop structure usually survives
the flare event.

\begin{figure}
\centering
\includegraphics[width=\hsize]{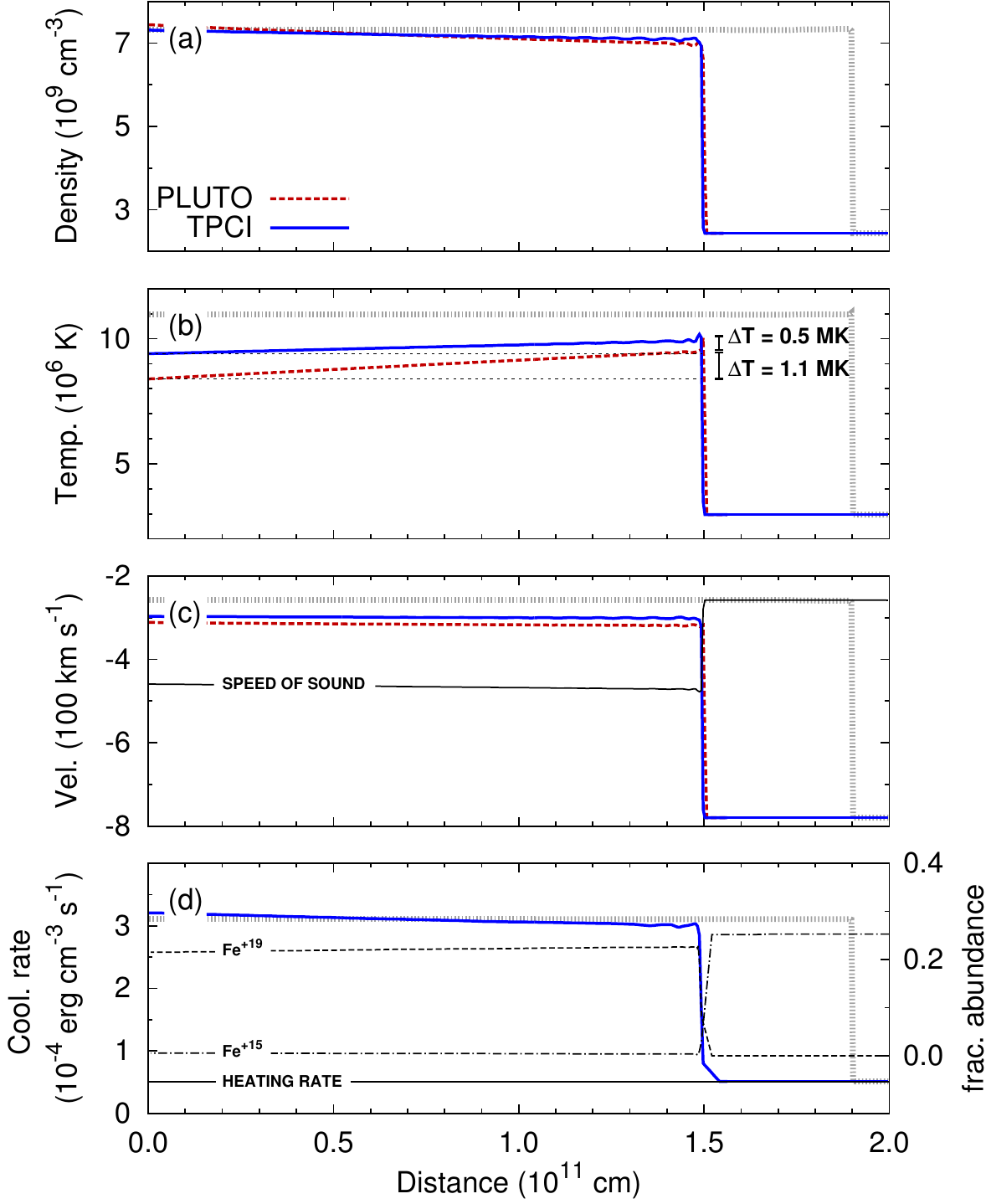}
   \caption{Hydrodynamic evaporation wave in a flaring coronal loop.
            The red dashed lines show a simulation with PLUTO
            assuming optically thin cooling.
            The blue lines depict the results of TPCI, and
            the initial conditions are given by the gray dotted lines.
            The density is shown in panel (a), the temperature in (b),
            the velocity in (c), and the cooling of TPCI is plotted
            in panel (d) together with the constant heating rate.
            Panel (d) shows the fractional abundance of the
            two iron ions with the strongest individual contribution to the total
            cooling rate.
            In panel (b) the temperature decrement in the 
            after-shock evolution is indicated.
            The sound speed is shown by the
            solid black line in panel (c).
            Both simulations show the same behavior, except that in the
            PLUTO simulation the cooling rate is twice as
            strong as in the TPCI simulation, which is due to optical-depth effects
            and small differences of the electron density.
           }
      \label{FigCoronalShock}
\end{figure}

For the simulation we went into the quasi-rest frame of the
propagating evaporation wave (see Fig.~\ref{FigCoronalShock} for the initial conditions).
The wave front is stable at the start of the simulation, but
the strength of the wave weakens as a result of cooling.
This reduces the propagation speed so that during the simulation
the wave front starts trailing toward the
left-hand boundary.
This proceeds faster in the PLUTO simulation
because of the stronger cooling rate.
Therefore, we do not compare contemporaneous states in both
simulations but states with the same displacement from the initial
shock position. There is no irradiation in this simulation.

In the test, coronal material flows in from the right-hand
side of the domain with a temperature of $3\times10^{6}$~K,
a hydrogen number density of $2.4\times10^9$~cm$^{-3}$,
and a velocity of $-780$~km\,s$^{-1}$.
The plasma is compressed to a density of $7.1\times10^9$~cm$^{-3}$
with a temperature of $10^{7}$~K and a velocity of $-300$~km\,s$^{-1}$.
The domain length was chosen deliberately longer than a typical
magnetic loop to emphasize the cooling of the dense gas
behind the shock.

To stabilize a coronal loop model, a constant mechanical
coronal heating rate has to be applied \citep{Rosner1978}.
This heating of unspecified source balances radiative and
conductive losses throughout the loop.
We applied a heating rate of
5.1/$9.8\times10^{-5}$~erg\,cm$^{-3}$\,s$^{-1}$
in the TPCI/PLUTO simulations, which exactly
equals the cooling rates in the pre-shock region
(see Fig.~\ref{FigCoronalShock}~(d)).
Advection of species can affect the ionization stage of metals
behind the shock, which in turn has a weak influence 
(factor~$<1.5$) on the value of the cooling function.
The increased density affects the cooling rate with a
factor of $\sim 3^2$,
however, which is why we neglected advection of species in this simulation.

Both simulations show the same behavior
(see Fig.~\ref{FigCoronalShock}).
The plasma is heated in the wave front and slowly cools down
in the post-shock region.
The cooling efficiency is a factor of two higher in the PLUTO
simulation than with TPCI. 
This has two reasons: 
first, the actual cooling rate in the TPCI simulation
is lower by a factor of 1.5 than in the optically thin case
because CLOUDY includes optical-depth
effects along the coronal loop (see Fig.~\ref{FigCoolingFuncs}).
For example, one of the strongest emission lines of the hot plasma
is a \ion{Fe}{XX} line at 12.9~\AA{}, which has an optical depth of 3.9
and does not escape freely.
Fe$^{+19}$ and Fe$^{+18}$ are the most abundant iron ions in the hot gas, while Fe$^{+16}$ and Fe$^{+15}$ dominate in the cooler gas.
Second, the PLUTO simulation computes the electron density
on a simple ``hydrogen-only'' approximation, which in this 
case overestimates the electron density by a factor of
1.2. Both factors combined result in the total difference of the
cooling rates ($1.2^2\times1.5=2.2$).

A comparison of the two simulations shows that TPCI can also be used for collisional plasmas,
especially if optical-depth effects are relevant.
One drawback is the 
increase in computational effort
by a factor of $10^{5}$ compared
to the much simpler, optically thin PLUTO simulation.
The strong increase in computational effort results from solving the
equilibrium state including all metals (see Sect.\ref{secDiscus}).
For future applications the gain in accuracy by using the radiative transfer
and microphysics
solver must be weighed against the increased computational effort.

\subsection{Formation of a Str\"omgren sphere}
\label{secRtype}

In a third setup, 
we used the interface to follow the heating of cool interstellar
hydrogen gas after the ignition of a strong central source.
This test problem demonstrates the limits of TPCI that are due to the underlying
assumption of ionization and chemical equilibrium.
The sudden impact of ionizing photons after ignition of the
source heats the surrounding
gas, and a supersonic ionization front develops (R-type).
The propagation of the ionization front is only restricted by the
rate of new ionizing photons emitted by the central source.
The rapid expansion comes to a rest after recombinations in
the ionized sphere balance the production rate of ionizing photons;
the resulting sphere is called a Str\"omgren sphere, and the 
radius is given by \citep[e.g.,][]{Osterbrock2006}:
\begin{equation}\label{eqStroemRad}
  R_{\mathrm{S}} = \left( \frac{3Q_{\mathrm{H}}}{4\pi n_0^2\alpha_{\mathrm{B}}} \right)^{1/3} \,.
\end{equation}
Here, $Q_{\mathrm{H}} = \int_{\nu_0}^{\infty} L_{\nu}/h\nu \,\mathrm{d}\nu$
is the number of ionizing photons emitted by the central source per second,
and $L_{\nu}$ is the stellar luminosity per unit frequency interval.
$n_0$ is the initial density and
$\alpha_{\mathrm{B}} = 2.59\times10^{-13}$cm$^3$s$^{-1}$ is the hydrogen
B recombination rate for a 10\,000~K plasma.
This rate includes recombination to all levels but the ground state
because recombination to the ground state produces
an ionizing photon that does not escape locally.
The R-type ionization front 
does not raise significant bulk motion because
it proceeds too fast through the interstellar medium
to cause significant acceleration.

We started with an initial constant density of
$10^4$~cm$^{-3}$, a temperature of 100~K, and zero velocity.
The emission rate of ionizing photons is $Q_{\mathrm{H}} = 10^{49}$~s$^{-1}$,
which is typical for a 50\,000~K hot O-star.
For the CLOUDY calls we used a spherical grid with the source in the center because
geometric dilution of the radiation field must be included.
Since no significant bulk flows develop throughout the simulation,
we simply used a Cartesian grid in PLUTO (irradiation from the center
in a spherical grid has not been implemented in TPCI yet), and
we also did not need to include the advection of species.
We only followed the
time-dependent heating of the initially cool interstellar gas.

Numerically, R-type fronts have been studied for example by \citet{Dale2007}.
The authors derived a simple phenomenological solution for the expansion of
the ionized region after ignition of the central source, which can be
used to compare the results of our code:
\begin{equation}\label{eqRfront}
  R(t) = R_{\mathrm{S}}\left( 1- \exp{(-n_0\alpha_{\mathrm{B}}t)} \right)^{1/3} \,.
\end{equation}
This formula is derived by noting that the radius increment per time
is restricted by the production rate of ionizing photons minus the
recombination rate inside the already ionized region.

\begin{figure}
\centering
\includegraphics[width=\hsize]{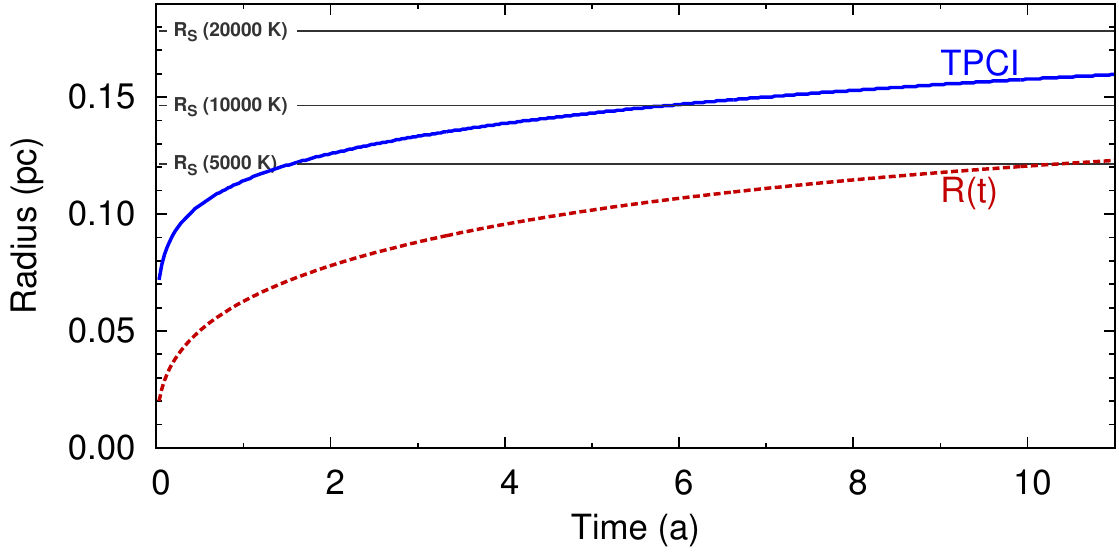}
   \caption{Expansion of an R-type ionization front.
            The TPCI simulation (blue solid) is compared
            with the theoretical evolution of Eq.~\ref{eqRfront}.
            Three Str\"omgren radii for different temperatures
            are indicated by labeled thin black lines.
            The expansion of the ionized region proceeds faster in the
            TPCI simulation because the equilibrium solutions of the
            photoionization solver at each integration step are not
            restricted by the time since ignition of the central source.
           }
      \label{FigRfront}
\end{figure}

Figure~\ref{FigRfront} shows the propagation of the R-type
ionization front in TPCI compared with the solution in Eq.~\ref{eqRfront}.
Especially the beginning of the simulation shows a faster expansion of
the ionized region than the theoretical prediction.
To understand the differences, we have to remember that CLOUDY
seeks equilibrium states.
Although the evolution seems to be similar to the theoretical
model, the simulation is based on a different physical mechanism.

\begin{figure}
\centering
\includegraphics[width=\hsize]{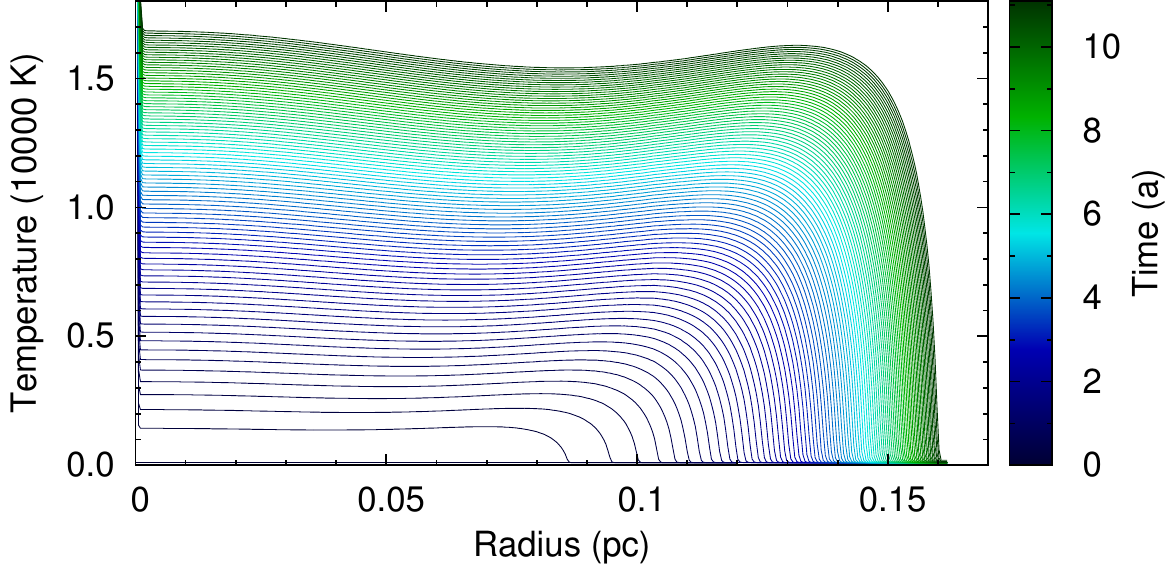}
   \caption{Temperature evolution during expansion of the
            R-type ionization front. The temperature versus
            radius is plotted at various times.
            The time step between consecutive lines is constant,
the time is referenced by the color scheme.
            The temperature increases from 1000~K to
            17\,000~K, being nearly constant throughout the ionized
            region.
            The expansion radius of the ionized region at a given
            time is marked by the steep temperature decline and 
            increases from 0.085 to 0.16~pc.
            The distance between consecutive lines decreases, indicating
            the approach to the final Str\"omgren radius.
           }
      \label{FigRfront_temp}
\end{figure}

Physically, the number of ionized hydrogen atoms cannot exceed the
number of emitted ionizing photons since the central source has been
switched on.
This is not the case in the TPCI simulation because the equilibrium photoionization
solver assumes that the source
already existed for an infinite amount of time.
The reason that we can nevertheless follow an expansion of the
ionized region is based on the fact that recombinations in the
cold gas are more frequent than in an almost fully ionized 10\,000~K hot
region.
Therefore, CLOUDY actually computes an equilibrium
Str\"omgren sphere at every time step,
but with the temperature structure that is passed from
PLUTO (see Fig.~\ref{FigRfront_temp}).
The Str\"omgren sphere increases because recombination becomes less
efficient with increasing temperature.
In Fig.~\ref{FigRfront} the Str\"omgren radii for 5000~K,
10\,000~K, and 20\,000~K are displayed.
At the times when the
TPCI simulation reaches the first two indicated Str\"omgren radii, the mean temperatures in the 
ionized region are 5600~K and 11\,400~K (compare Figs.~\ref{FigRfront}~and~\ref{FigRfront_temp}).
The photoionization solver does not assume the simple B recombination
rate as in Eq.~\ref{eqStroemRad}, but allows photons to escape based on 
the optical depth. Hence, recombination in the simulation is more
efficient, which causes the mean temperatures to be slightly higher at the
given Str\"omgren radii.

This simulation clearly shows that the user must be
aware of the restrictions of TPCI and what is actually
simulated.
The interface cannot be used to correctly simulate the
expansion of an ionized region after ignition of a central source
because this is an effect of a time-dependent radiation field.
It could, however, be used to follow the subsequent expansion
of the hot gas in the Str\"omgren sphere into the surrounding
cool interstellar gas under unchanged irradiation by the O-star
because this can be approximated by equilibrium states.

\subsection{Simulation of the atmosphere of HD\,209458\,b}
\label{secHJatmos}

The intention of the design of TPCI was simulating hot-Jupiter atmospheres, and here
we present a first test case by solving a pure hydrogen atmosphere of HD\,209458\,b as our fourth test case.
The expanded atmosphere around HD\,209458\,b was discovered by
\citet{Vidal2003}, and several studies investigated the system numerically
\citep[e.g.,][]{Yelle2004, Tian2005, Garcia2007, Penz2008-2, Murray2009, Koskinen2013}.
These studies provide a solid basis for comparing the results from TPCI.

For the simulation, we made use of two more modules of
PLUTO: gravitational acceleration and thermal conductivity.

\subsubsection{Gravity}
\label{secGrav}

The simulation of the escaping hot-Jupiter atmosphere incorporates
the gravitational and centrifugal acceleration in the two-body system,
also referred to as tidal forces.
With the planet in the center, we derive for the effective potential
\begin{equation}\label{eqGravPot}
  \Phi (r) = -\frac{GM_{\mathrm{pl}}}{r}
             -\frac{GM_{\mathrm{st}}}{a-r}
             - \frac{1}{2}\omega^2(l_{\mathrm{cm}}-r)^2\,.
\end{equation}
Here $r$ denotes the radial distance on the axis toward the star,
$M_{\mathrm{pl}}$ and  $M_{\mathrm{st}}$ are the planetary and stellar mass,
$a$ is the semi-major axis, $l_{\mathrm{cm}}$ is the distance to the
center of mass, and $G$ is the gravitational constant.
The angular frequency $\omega$ is given by Kepler's third law
\begin{equation}\label{eqPlanetPeriod}
  \left(\frac{2\pi}{T}\right)^2 = \omega^2 = \frac{G(M_{\mathrm{st}}+M_{\mathrm{pl}})}{a^3} \,.
\end{equation}
In PLUTO we include the acceleration:
\begin{equation}\label{eqGravAccel}
  a_{\mathrm{G}} = - \frac{\partial \Phi}{\partial r}
           = -\frac{GM_{\mathrm{pl}}}{r^2}
             +\frac{GM_{\mathrm{st}}}{(a-r)^2} 
             -\frac{G(M_{\mathrm{st}}+M_{\mathrm{pl}})}{a^3}(l_{\mathrm{cm}}-r) \,.
\end{equation}
By plotting the equation with the values of the system HD\,209458
\citep{Wright2011}, we find that 
on the star-planet axis material can be bound to the planet up to
4.18~R$_{\mathrm{pl}}$,
which is the size of the planet's Roche lobe on this axis.

\subsubsection{Thermal conductivity}

\begin{figure}
\centering
\includegraphics[width=\hsize]{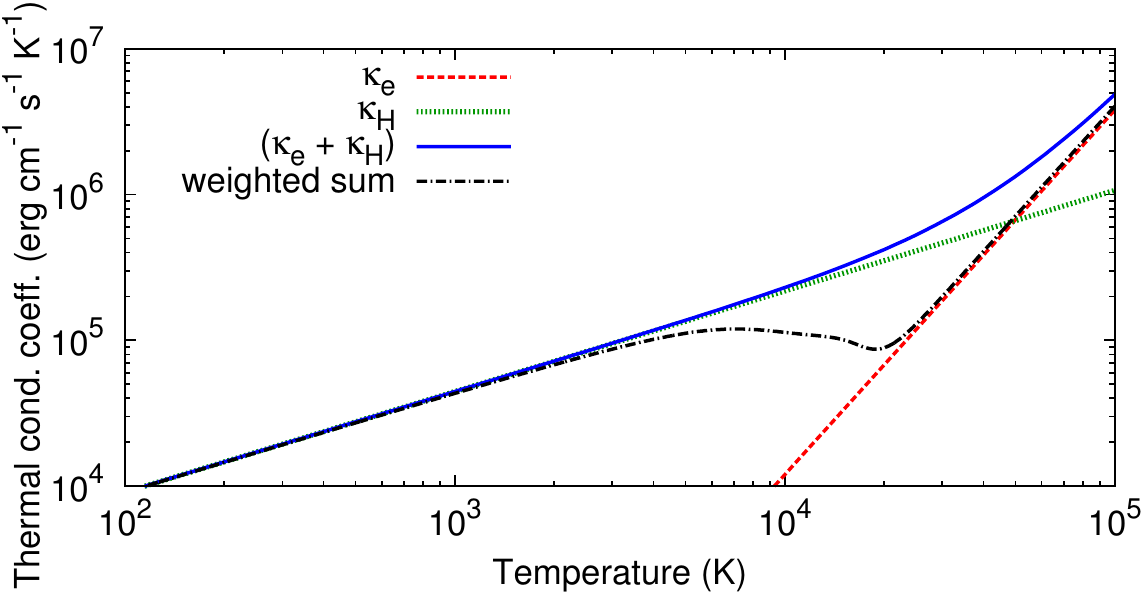}
   \caption{Thermal conductivity coefficient.
            The coefficient for electron conductivity
            is depicted by the red dashed line,
            for neutral hydrogen by the green
            dotted line,
            and the blue solid line is the sum.
            The black dash-dotted line shows the more
            accurate weighted sum (Eq.~\ref{eqTCweight})
            derived with the actual densities of neutral hydrogen and
            electrons in a CLOUDY
            simulation without irradiation.
            In the temperature region below 10\,000~K
            the simple sum is accurate to within a 
            factor of two.
           }
      \label{FigTCkappa}
\end{figure}

We used a simplified approach to include thermal
conductivity
by adding the conductivity of an
electron proton plasma and
of a neutral hydrogen gas.
The first dominates for temperatures in excess of
50\,000~K, the latter below this.

The following coefficients are taken from \citet{Banks1973}.
The heat flux is given by
\begin{equation}\label{eqCondFlux}
  \mathbf{F_c} = -\kappa \,\nabla\!\left( T \right) \,.
\end{equation} 
For highly ionized gases, thermal conductivity is dominated by
electron-ion collisions \citep{Spitzer1978}. The 
conductivity coefficient is given by
\begin{equation}\label{eqEcond}
  \kappa_{\mathrm{ei}} =
  1.2\times10^{-6} T_{\mathrm{e}}^{5/2} \,.
\end{equation} 
The thermal conductivity coefficient of neutral hydrogen is parameterized as
\begin{equation}\label{eqHcond}
  \kappa_{\mathrm{HH}} = 379 \; T^{0.69} \,,
\end{equation}
and the total coefficient is given by the weighted sum
\begin{equation}\label{eqTCweight}
  \kappa = \frac{1}{n} \sum n_\mathrm{s} \kappa_\mathrm{s}
         = \frac{1}{n} \left( n_\mathrm{e} \kappa_\mathrm{e}
            + n_\mathrm{H} \kappa_\mathrm{H}\right) \,.
\end{equation}
In our simulations we used the direct sum of the two coefficients:
\begin{equation}\label{eqTCsum}
  \kappa = \kappa_\mathrm{e} + \kappa_\mathrm{H} \,.
\end{equation}
In the relevant temperature range below 10\,000~K
the approximation is correct to within a factor of two (see Fig.~\ref{FigTCkappa}).
Since we do not expect thermal conductivity to have a significant
impact on the results \citep{Garcia2007},
the procedure is sufficient to verify this claim in our simulation.

\subsubsection{Simulation}
\label{secHDTWO}

\begin{figure}
\centering
\includegraphics[width=\hsize]{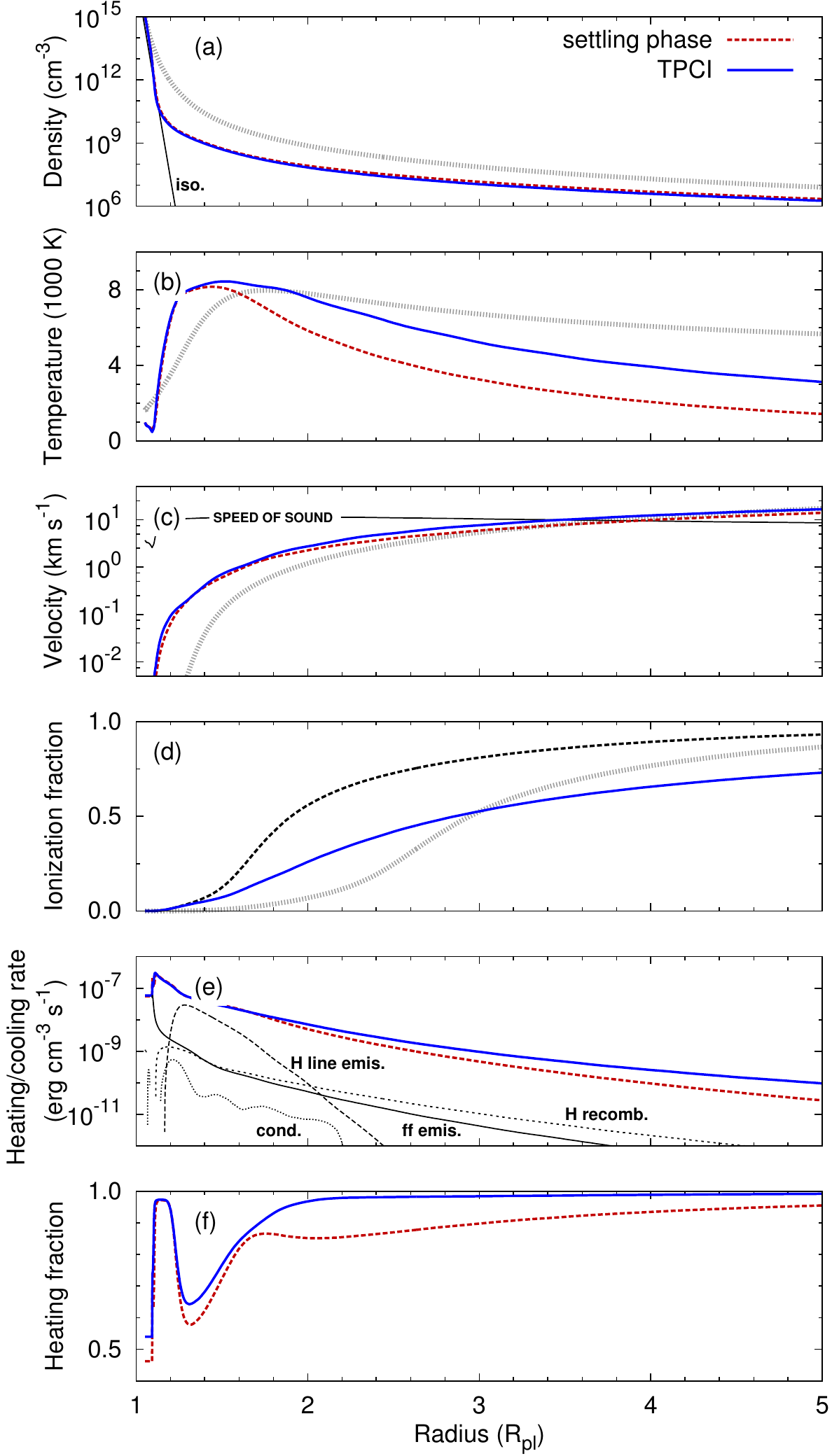}
   \caption{Evaporating hydrogen atmosphere of HD\,209458\,b.
            Density (a), temperature (b), velocity (c),
            ionization fraction (d), heating and cooling rate (e),
            and the heating fraction (f) are plotted versus
            the planetary radius.
            The red dashed lines depict the initial settling phase, the blue solid lines the final TPCI simulation
            including advection of species.
            The initial conditions are given by the gray dotted lines.
            Panel (a) also shows the density structure of an
            isothermal atmosphere with 500~K (black solid line).
            In panel (c) the speed of sound is indicated
            (black solid line),  and in panel (e) the three
            main cooling agents in the atmosphere are shown
            by black thin lines. ``H line emis.'' is cooling due
            to line emission from hydrogen atoms (mainly 
            Ly$\alpha$), ``ff emis.'' is cooling by free-free
            emission of electrons, and ``H recomb.'' is recombination
            cooling of hydrogen. 
            Additionally, thermal conductive cooling (cond.) is
            shown by a black dotted line.
            The atmosphere expands in a transonic flow.
            A small amount of the absorbed energy ($\sim 11$\%) is
            lost through free-free and Ly$\alpha$ emission.
           }
      \label{FigPenzHD209}
\end{figure}

We used a pure hydrogen atmosphere, irradiated from the top
by the host star, and simulated the escape
at the substellar point.
With a stretched grid the cell size is increased from 
0.0005~R$_{\mathrm{pl}}$ in the lower atmosphere to 
0.087~R$_{\mathrm{pl}}$ at the top.
The temperature in the lower atmospheres was fixed at the planet facing boundary to
1000~K, which is approximately the average dayside brightness temperature of HD\,209458\,b \citep{Crossfield2012}.
In the simulation this was implemented by fixing the lower boundary density
to $10^{15}$~cm$^{-3}$ and the 
pressure to 140~dyn\,cm$^{-2}$ , which is equivalent to 0.14~mbar.
The gas is heated by ionizing radiation with a solar minimum
EUV spectral energy distribution \citep{Woods2002}, which is a
good approximation for an inactive G0-type
dwarf \citep{Sanz2011,Montes2001}. 
In the spectral range below 912~\AA{} (XUV), the planetary atmosphere at the distance of
0.047~AU from the host star is irradiated with a 
flux of $F_{\mathrm{XUV}} = 1315$~erg\,cm$^{-2}$\,s$^{-1}$.
The derived value is similar to previous estimates, for instance, 
1800~erg\,cm$^{-2}$\,s$^{-1}$ in \citet{Koskinen2013}.
For this inactive host star \citep{Sanz2011}, X-rays ($<$\,100~\AA{}) contribute only 3\% of the energy in the XUV spectral range. In contrast to active host stars, X-rays have no strong impact on the mass loss of HD\,209458\,b.
The initial conditions of the simulation are shown together with the
results in Fig.~\ref{FigPenzHD209}.
A comparable setup can be found in the simulations of
\citet{Penz2008-2}, \citet{Murray2009} or \citet{Koskinen2013}.

In the expanding atmosphere neutral hydrogen
is transported from the lower to the upper atmosphere, which
means that advection of
species is essential for the results in this simulation.
To save computational time, we ran the simulation without the
advection of species until a steady-state planetary wind was found and 
all shock waves had subsided.
The resulting atmosphere was used as initial atmosphere for the final simulation
including the advection of species.
The convergence of the advective scheme does not proceed beyond
an advection length of 0.27~R$_{\mathrm{pl}}$.
This is comparable to the grid resolution at the top of the
atmosphere, 
but much higher than the resolution in the lower atmosphere,
so that effects due to the advection of species are initially only 
resolved in the upper atmosphere.
However, below 1.05~R$_{\mathrm{pl}}$ velocities are low and the advective 
timescale is longer than the hydrogen recombination timescale.
Advection of species can safely be neglected below this level.
Furthermore, we manually reduced the advection length to 0.01~R$_{\mathrm{pl}}$
and did not find significant changes in the solution.
We therefore conclude that the advection of neutral hydrogen into the
upper atmosphere is resolved throughout this simulation.

\subsubsection{Results}

The atmosphere expands in a steady-state transonic flow that is directed from the
hot Jupiter toward the host star.
We find a steep density and temperature gradient within the
first 0.2~R$_{\mathrm{pl}}$ (see Fig.~\ref{FigPenzHD209}).
At 1.08~R$_{\mathrm{pl}}$ the atmosphere shows a temperature
minimum with only 500~K at a pressure of 0.7~$\mu$bar; the density stratification below this
point differs only slightly from an isothermal atmosphere
with the same temperature (see Fig.~\ref{FigPenzHD209} (a)).
EUV radiation does not penetrate to this depth, but is mainly absorbed
from 1.1 to 1.2~R$_{\mathrm{pl}}$, where
the temperature of the atmosphere strongly increases.
In this pure hydrogen atmosphere free-free emission
is the main radiative cooling agent below 1.08~R$_{\mathrm{pl}}$,
reemitting about 50\% of the total heat input, up to 70\% of which is provided
by hydrogen line absorption at this depth.
Hydrogen line emission, mainly Ly$\alpha$ radiation,
dominates the radiative cooling from 1.08 to 2.0~$R_{\mathrm{pl}}$,
and recombination of hydrogen is the most
efficient radiative cooling agent at higher altitudes.

Averaged over the atmosphere above 
the temperature minimum,
89\% of the heat input is used for
heating and acceleration of atmospheric gas, and eventually for lifting the material out of the gravitational potential well of the planet.
The total energy gain is divided as follows:
77\% of the radiative energy input is converted to gravitational potential energy, 17\% to kinetic and 6\% to internal energy.
Note that we define the heating fraction as
\begin{equation}\label{eqHeatFrac}
  f_\mathrm{H} = \frac{G_{\mathrm{R}} - L_{\mathrm{R}}}{G_{\mathrm{R}}}\, ,
\end{equation}
here $G_{\mathrm{R}}$ is the actual heating produced by the absorption of 
radiation, not the energy of the absorbed radiation (see Eq.~\ref{eqIonizHeatRate}).
In our hydrogen-only simulation,
only free-free and Ly$\alpha$ emission decrease
the radiative heating fraction significantly (see Fig.~\ref{FigPenzHD209} (f)).
The fraction approaches 1.0 in the upper atmosphere;
however, emission by helium or metals is not included in this simulation.
Especially line absorption and emission by metals can significantly affect the heating and cooling rates in the lower atmosphere.

Replacing the denominator in Eq.\ref{eqHeatFrac} with the absorbed radiative energy
gives the net heating efficiency, which is often used
in hydrodynamic simulations of hot-Jupiter atmospheres to account for
the fraction of energy that goes into ionizing hydrogen and also for
not specifically included cooling agents.
In this particular simulation, stellar emission is absorbed from 50 to 912~\AA{} by the atmosphere (more than 50\% absorption). This absorption produces 92\% of the total heating rate; the remaining 8\% are caused by hydrogen line absorption, mainly Ly$\alpha$ absorption. The heating efficiency in the XUV range is 0.71.

Thermal conduction does not play a significant role in the atmosphere since
it is more than two orders of magnitude smaller than the main cooling
agents at every point throughout the atmosphere (see Fig.~\ref{FigPenzHD209} (e)).
Comparing the structure from the initial settling phase without advection of
species with our final results shows that the advection of neutral hydrogen into
the upper atmosphere increases the average temperature from 3000 to 5500~K.
Advecting neutrals into the otherwise
highly ionized region
increases the heating rate
because more ionization processes occur.
This also boosts the heating fraction, as can be
seen in Fig.~\ref{FigPenzHD209}.
The final velocity exceeds the sonic speed of 10~km$^{-1}$.

Our results are in excellent
agreement with the C3 model of \citet{Koskinen2013} for the
atmospheric evaporation at the substellar point.
Assuming a spherically symmetric atmosphere, 
we derive a mass-loss rate of
$\dot{M} = 4\pi r^2\rho v = 1.4\times10^{11}$~g\,s$^{-1}$, 
which is equal to 0.0023~M$_{\mathrm{Jup}}$\,Ga$^{-1}$.
This value is an upper limit since we simulated
the substellar point with the highest mass-loss rate.
For the surface averaged mass loss, a factor of 1/4 was used
by \citeauthor{Koskinen2013}, which places our result just slightly below their
values, ranging from $4 - 6.5\times10^{10}$~g\,s$^{-1}$.
Within 5\% accuracy, advection of species does not change
the mass-loss rate because a slightly higher velocity
in the advective simulation
is canceled by a lower density.

\citet{Murray2009} were the first authors to include a model for Ly$\alpha$ cooling
in the simulation of escaping planetary atmospheres.
Their spatial distribution and total cooling rate compares well with our results.
Ly$\alpha$ cooling leads to a lower peak temperature in the atmosphere
and reduces the mass-loss rate \citep{Koskinen2013}.
Note, however, that deeper in the atmosphere the density
is high enough that
collisional de-excitation dominates and absorption of
Ly$\alpha$ radiation is a heating agent. 
About one quarter of the total cooling through Ly$\alpha$ radiation is counterbalanced by
Ly$\alpha$ heating in the lower atmosphere.
On the other hand, \citeauthor{Murray2009} used a gray absorption scheme,
which leads to a mass-loss rate lower by a factor of four \citep{Koskinen2013}.
Such differences clearly prove the advantages of the 
detailed photoionization solver included in TPCI.
Neglecting any significant cooling or heating agent
affects the mass-loss rate,
which crucially depends on the available energy.
This becomes even more important when
metals are included in the atmosphere and an increasing number of
processes must be modeled.

Although the mass-loss rate does not change when
the advection of species is included in the simulation,
the advection of neutral hydrogen into the upper atmosphere
is important for analyzing observational results.
The Ly$\alpha$ absorption signal of the planet during transit
does change significantly because of the higher neutral hydrogen
column density. The high velocities of the expanding atmosphere are crucial to
transport the neutral hydrogen into the upper atmosphere.

\section{Discussion and conclusion}
\label{secDiscus}

We have presented TPCI, which is an
interface between the MHD code PLUTO and 
CLOUDY, an equilibrium photoionization solver.
TPCI enables simulations of hydrodynamic steady-state and 
slowly evolving flows, including multiple physical effects such
as gravity and thermal conduction. It simultaneously solves the equilibrium state
of a gas or plasma under strong irradiation.
The code is valid from 
cold molecular regions to fully ionized hot
plasmas and computes the radiative transfer along 1D rays.

The presented test simulations prove the validity of the combined
simulations.
In the introduction we have established a number of requirements for TPCI. The interface has proven its capabilities in solving the ionization and microphysical state of gases for a wide range of densities and temperatures. The second test problem (see Sect.~\ref{secCoronalFlare}) demonstrates the simulation of gases including metals and shows effects of X-ray radiative transfer. Multidimensional simulations or magnetic fields have not been studied in our verification of the 1D scheme.
In general, we find that TPCI runs stably and reliably.
The combination of the adaptable hydrodynamics code with a 
versatile
radiative transfer and microphysics solver is generally
applicable to photoevaporative flows and can for example be used to study
the dynamics of \ion{H}{II} regions or the evaporation of protoplanetary disks.
Furthermore, CLOUDY is a powerful tool for spectral analysis,
and observed spectra can be directly
compared with results from the simulations.

One drawback in the use of TPCI is the computational effort
of the detailed photoionization solver, which 
usually consumes almost all of the computational time.
For the simulation of the test problems we used a few days
to up to two months on a single 3.1~GHz processor.
One-dimensional simulations are serial, so that simulations
do not perform faster on clusters.
In the search for hydrodynamic steady-state situations, the
best call for reducing the computational effort is to
find good initial conditions to speed up the convergence
and to include more complex physics step by step from
pre-converged simpler solutions.

\begin{figure}
\centering
\includegraphics[width=\hsize]{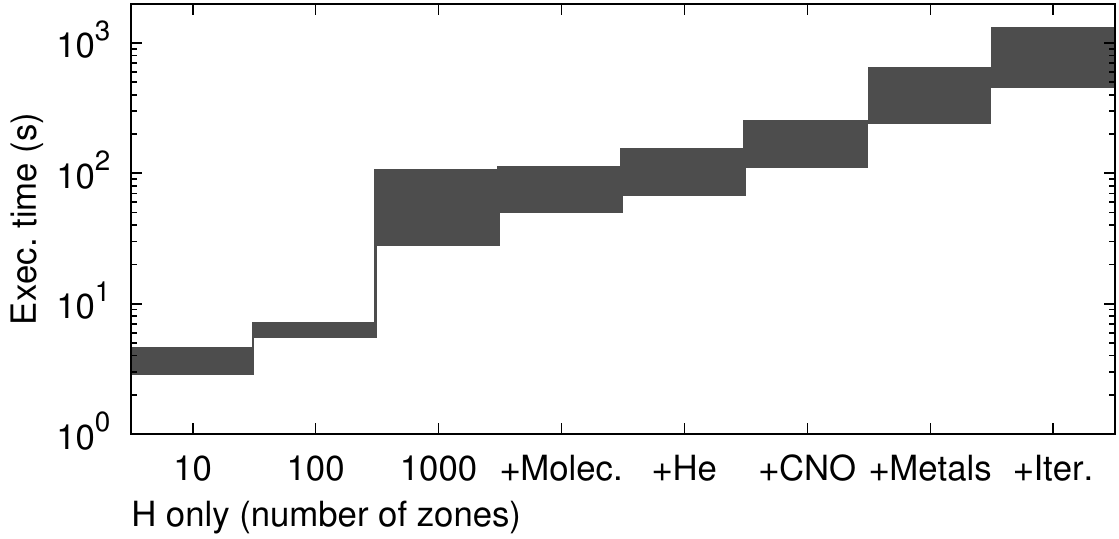}
   \caption{Execution times of CLOUDY runs on local cpu
            with setups based on the hot-Jupiter simulation.
            Simulation times vary in the shaded range based
            on density and temperature.
            Each step increases the complexity of the simulation,
            starting with a simulation including only hydrogen
            and 10 zones in CLOUDY. In this case the execution time
            is dominated by the initialization of CLOUDY.
            The next steps are with 100 and 1000 zones. 
            We then include step by step molecules, helium,
            carbon, nitrogen, and oxygen, and all 30 lightest
            elements.
            Finally, we iterate twice in CLOUDY over the
            complete domain to correct optical-depth effects.
           }
      \label{FigCloudyExecTimes}
\end{figure}

Figure~\ref{FigCloudyExecTimes} shows execution times in 
single calls to the photoionization solver.
The setup for the calls is based on the simulation of the hot-Jupiter atmosphere, but we used constant hydrogen number densities
ranging from $10^5$ to $10^{15}$~cm$^{-3}$ and constant
temperatures ranging from 1000 to 10\,000~K.
The actual execution times vary depending on the parameters.
In a typical simulation of a hot-Jupiter atmosphere CLOUDY
needs about 1000~zones for the adaptively adjusted grid,
which means that execution times were about 100~s, but 
two iterations are needed to correctly compute the
optical depth for the escape probability formalism.
The settling phase of the simulation took about 10\,000
calls to the photoionization solver, which explains the
simulation time of about one month.
For the advection of species, at least 30 iterations have to be made
in the CLOUDY internal steady-state solver,
but when starting from the previously converged simulation, about
200 calls to the photoionization solver suffice, which adds up to
one day.
In the future, the abundance of metals can be linearly increased during the simulation
to prevent strong oscillation in the atmosphere and, thus,
limit the CLOUDY calls to a minimum.

The convergence of the advection of species in the CLOUDY internal
steady-state solver can cause problems in certain situations.
The result is a solution where effects of the advection of species
are only resolved on scales larger than the advection length.
One possible solution is to manually reduce the
advection length to the needed accuracy, which
increases the computational effort, however, because more iterations in CLOUDY
are needed.
Furthermore, advection can only be included in 1D simulations
with flows toward the irradiating source.
A possible solution would be to include passive
scalar fields in PLUTO for every species
and solve the advection by explicitly including source and sink terms,
which could be retrieved from each call to the photoionization solver.
This complicates the interface, however and necessitates more essential changes
of the CLOUDY source code because the number densities of
each species would have to be passed to CLOUDY and source and
sink terms
must be retrieved.
Species include all ionization stages of the 30 lightest elements plus
all included molecules.

The simulation of atmospheric evaporation in the
well-studied hot Jupiter HD\,209458\,b 
demonstrates the capabilities of TPCI.
The simulation of the hydrogen-only atmosphere shows
significant differences compared to
\citet{Murray2009} because the authors used a gray absorption scheme, 
or compared to \citet{Penz2008-2}, who neglected Ly$\alpha$ cooling.
Our atmosphere shows an excellent agreement with the corresponding
model of \citet{Koskinen2013}.
We resolved a temperature minimum in the lower atmosphere
because our simulation continues deeper into the atmosphere.
Furthermore, we used a more comprehensive microphysics solver, which reveals
significant cooling due to free-free emission in the lower
atmosphere, which has previously been
underestimated \citep{Murray2009}.

The increased accuracy in solving the absorption and emission
processes is a clear advantage, and by using the
well-tested CLOUDY code, we did not 
neglect any significant processes in our model.
The total change in the mass-loss rate compared to the model of \citet{Koskinen2013}
is lower than a factor of two in this pure hydrogen atmosphere, but
larger differences can be expected when metals will be included.
In the future, we will extend this initial test case to include molecules and
metals and will investigate the effects of strong X-ray irradiation
in the atmospheres of planets around active host stars.

\begin{acknowledgements}
We thank G. J. Ferland for a discussion of the problem.
This research has made use of the Exoplanet Orbit Database
and the Exoplanet Data Explorer at exoplanets.org.
M.S. acknowledges support from the
\emph{Deut\-sches Zen\-trum für Luft- und Raum\-fahrt, DLR\/}
under the project 50OR1105 and the 
\emph{Deut\-sche For\-schungs\-ge\-mein\-schaft, DFG\/} 
RTG-1351.
\end{acknowledgements}


\bibliographystyle{aa}
\setlength{\bibsep}{0.0pt}
\bibliography{salz_TPCI}

\end{document}